# An explanation of unexpected *Hoxd* expressions in mutant mice

**Spyros Papageorgiou***

Institute of Biology, NCSR 'Demokritos'

Athens, Greece

**Running Head:** Explanation of unexpected Hoxd expressions

***Corresponding author:**

**Spyros Papageorgiou**

**Institute of Biology, NCSR 'Demokritos', Athens, Greece**

**Tel: (30)-210-8954920, FAX: (30)-210-7750001**

**Email: spapage@bio.demokritos.gr**

**Abstract**

The *Hox* gene collinearity enigma has often been approached using models based on biomolecular mechanisms. The 'biophysical model', is an alternative approach, speculating that collinearity is caused by physical forces pulling the *Hox* clusters from a territory where they are inactive to a distinct spatial domain where they are activated in a step by step manner.

  *Hox* gene translocations have recently been observed in support of the biophysical model. Furthermore, genetic engineering experiments, performed in embryonic mice, gave rise to some unexpected mutant expressions that biomolecular models could not predict. In several cases when anterior *Hoxd* genes are deleted, the expression of the genes whose expression is probed in the mutants are 'impossible to anticipate'. On the contrary, the biophysical model offers convincing explanation.

  All these experimental results support the idea of physical forces being responsible for *Hox* gene collinearity. In order to test the validity of the various models further, certain experiment involving gene deletions are proposed. The biophysical and biomolecular models predict different results for these experiments, hence the expected outcome will confirm or question the validity of these models.

**Introduction**

In 1978 E.B. Lewis discovered the phenomenon of *Hox* gene collinearity [1]. According to his observations in *Drosophila,* the *Hox* genes are activated following the sequential spatial order anterior-to-posterior on the embryonic primary axis. Surprisingly, these genes are located on the chromosome following the same order. The evolutionary origin of this collinearity has been extensively studied [2,3,4].

  The last decade genetic engineering experiments have illuminated several features of the still enigmatic phenomenon of expression collinearity of the clustered *Hox* genes. In a series of experiments on the *HoxD* cluster in embryonic mice, members of this cluster were deleted. In some cases the deleted genes were anterior to a probe gene whose expression was analyzed in detail. In particular, the expression of *Hoxd11* in *wild type* embryos was compared to that in mutant embryos in which the anterior region [*Hoxd8-Hoxd10*] was deleted (Fig.1). The mutant expression shows *wild type* spatial distribution [5]. In contrast, when the region [i-*Hoxd8-Hoxd10*] was deleted, the expression of *Hoxd11* was dramatically extended anteriorily. In deletion del(i,10), adjacent to del (8,10), the intergenic DNA fragment " i" was included (Fig.1). " i" is located between *Hoxd4* and *Hoxd8*. The results were 'impossible to anticipate' [5].

As reported recently in a 3D *in vivo* analysis of conformational chromatin modifications during *Hox* cluster activation, *Hox* genes move step by step from a compartment where the cluster is inactive to a spatially distinct domain where *Hox* genes are transcriptionally active [6].

Combining the above experiments, it was found that the mutant *Hoxd11* with deletion del(8,10) was not ectopically expressed in the anterior trunk of the mouse embryo and it was not associated with the active part of the cluster. In contrast, the mutant *Hoxd11* with deletion del(i,10) was ectopically expressed in this anterior trunk and it was strongly associated with the active part of the *HoxD* cluster [6]. It is a challenge to understand the above combined experimental results.

**Discussion**

The established mechanisms proposed to account for the collinearity phenomena of *Hox* clusters are based solely on biomolecular processes [5,7]. However, these mechanisms have failed to describe all data and it seems that some new factors beyond biochemistry are responsible. With this in mind a different approach was proposed, more than a decade ago, that incorporates physical principles [8-13]. (Since this note is essentially based on references 12 and 13, in the following these references are denoted by **A** and **B** respectively). The conceptual motivation for this alternative approach was the multiscale organization of *Hox* gene collinearity. In order to deal with multiscale phenomena, Systems Biology seeks the involvement of mechanisms from other disciplines like Physics and Mathematics (see details in **B**).

According to the biophysical model for *Hox* gene collinearity during the early embryonic stages, a physical force is generated which pulls and gradually translocates the *Hox* genes from the Chromatin Territory (CT) toward the Interchromosome Domain (ICD) where the transcription factories are located [14] (Fig.2). In CT the *Hox* cluster remains inactive whereas in ICD gene activation is initiated. The case of a simplified Coulomb force **F** was explicitly studied where

$$F = P*N.$$

Here N is the total negative electric charge of the *Hox* cluster and P the positive charges of the attracting environment located opposite the 3' (anterior) end of the cluster. The P-values form a gradient along the anterior-posterior axis of the embryo and depend on the cell position on the embryo axis with the gradient peak at the posterior tip [15].

A mechanical analogue of *Hox* cluster activation is illuminating: the decondensation and translocation of *Hox* clusters can be thought as resembling an elastic spring. The spring can expand and the spring elongation is proportional to the pulling force **F** [**A,B**]. The anterior end of the spring is loose and can be pulled by a force while the posterior end of the spring is fixed inside the chromosome territory. The linear dependence of the spring elongation on the pulling force F is, of course, an oversimplification.

**Derivation of the 'impossible to anticipate' results**

The biophysical model can explain the genetic engineering data at the early developmental stages [**A,B**]. Additionally, with limited refinements, the model can also explain the unexpected findings [5,6] observed at later stages.

An anterior DNA deletion affects a probe gene expression following two consecutive steps: a) the deletion D causes a reduction of N due to the removal of the charges of this DNA region. Hence the normal pulling force **F** will be reduced to **Fc**. b) a consequence of the weaker force **Fc** is that the extruded DNA fiber is shorter than the *wild type* extruded fiber length (Fig.2). Suppose that the extruded length of the *wt* fiber from the anterior end of the cluster to the probe gene is L. The anterior deletion D will cause a shortening of this length to (L-D). Schematically this is depicted in Fig.1. Consider now E the extruded fiber of the mutant probe gene (Fig.2). E and (L-D) are not necessarily equal because the local electric charges and the DNA elastic properties differ from place to place along the chromatin fiber. E depends on these local properties of the DNA fiber since this fiber is strongly inhomogeneous.

There are three possible cases for mutant *Hoxd* expressions after an anterior gene deletion D (Fig.2).

1. (L-D) = E. The deleted region D causes an equal length reduction of the extruded gene fiber. (The ideal case of uniformly distributed negative charges and elastic properties along the whole DNA fiber could belong in this class of deletions). For the gene whose expression is probed, the position in relation with the interface CT/ICD does not change (Fig.2b).

2. (L-D) < E. The mutant extruded fiber E is longer than (L-D) and the gene whose expression is probed shifts inside ICD. A shorter fiber extrusion may suffice for the expression of this mutant gene (Fig.2c). According to the Coulomb force $F = P*N$ this shortening is the result of a weaker force F. For given N this can be achieved by reducing P. The reduction of P leads to an anteriorization of this mutant gene expression as elaborated in detail elsewhere [**A,B**].

3. (L-D) > E. In this case, the mutant gene remains inside CT and no activation of this gene will be observed (Fig.2d). Activation of the mutant gene is possible only if G is translocated to ICD. This can occur by a P increase (posteriorization) [**A,B**].

It is interesting to look for actual manifestations of the above cases. Although the biophysical model applies to the early developmental stages (up to about E9.5) the conclusions can be extended to later stages for a comparison with the existing experimental data.

In Case 1 the shortening of the extruded fiber equals the length of the deleted region so that the position of the mutant gene remains invariant (Fig.2b). This possibility can explain the observation of Tschopp *et al.* [5]: the mutant *Hoxd11* expression after deletion del(8,10) is comparable with the *wt Hoxd11* expression.

The observed ectopic anteriorization of the *Hoxd11* expression after deletion del(i,10) can belong to Case 2 (Fig.2c): the extruded fiber exceeds the length (L-D) and the mutant gene *Hoxd11* moves inside ICD. Therefore a retreat toward interface CT/ICD is permissible and an anteriorization of *Hoxd11* may occur for deletion del(i,10). This anteriorization was observed by Tschopp *et al.* [5]. Note that a 'dramatic gain' of the mutant *Hoxd11* expression was noticed and this is understandable since *Hoxd11* shifts toward the interface ICD/CT where the gene activation is stronger in the area of the transcription factory [**A,B**].

Case 3 can explain the observed expressions of *Hoxd13* after anterior deletions (see Fig.5 G, H, I in Ref.5): deletion del(10,12) leaves the expression of *Hoxd13* unchanged compared to the *wt* expression. This behavior is reminiscent of the mutant *Hoxd11* expression after deletion del(8,10) (see above). In contrast, the longer deletion del(9,12) leads to a strong suppression of *Hoxd13*. Note that *Hoxd13* is the most posterior gene of the cluster so that further posteriorization is impossible. Therefore, after deletion del(9,12) (Fig.2d) and according to the biophysical model, *Hoxd13* remains inside the CT area where it cannot be activated. This is in agreement with the observed strong suppression of *Hoxd13* expression [5].

On one hand, the above analysis explains the *prima facie* unexpected expressions of mutant Hoxd genes. On the other hand, it provides evidence supporting the hypothesis that physical forces cause the collinearity of *Hox* gene expressions.

**Further predictions and conclusions**

1. Consider Cases 1 and 2 for deletions del(8,10) and del(i,10) and the corresponding extruded fibers E(8,10), E(i,10) for the mutant *Hoxd11*. Consider furthermore E(i) the extruded fiber where only the intergenic region (i) is deleted. The biophysical model predicts that for del(i) the mutant *Hoxd11* expression <u>should be anteriorized.</u> This can be easily proved since del(i,10) = del(8,10) + del(i) and E(i,10) = E(8,10) + E(i). According to Case 1 for del(8,10) and Case 2 for del(i,10)

$$E(i,10) = E(8,10) + E(i) > [L - \text{del}(8,10)] + [L - \text{del}(i)]$$

E(8,10) = L – del(8,10) as derived from Case 1. Therefore,

$$E(i) > L - \text{del}(i)$$

The extrusion E(i) exceeds L-del(i) and the corresponding mutant *Hoxd11* moves inside ICD and its expression can be anteriorized (Fig.2c). This is a concrete biophysical model prediction worth testing. Such a behavior is not expected according to the established biomolecular models [5].

2. In analogy to the above mutant *Hoxd11* expressions, it is straightforward to formulate another prediction: comparing the mutant *Hoxd13* expressions after deletions del(10,12) and del(9,12), the biophysical model predicts that the mutant *Hoxd13* expression will be down regulated after deletion of only the *Hoxd9* region. It would be interesting if this prediction were also tested.

3. Posterior deletions are simpler to deal with and only different degrees of posteriorizations for the mutant genes are expected from such deletions [**A,B**]. *Hoxd* duplications are treated as indicated in [**A**] and the existing results agree with the biophysical model predictions.

4. The biophysical model answers the question whether the physical separation of active from non-active *Hox* genes 'underlies collinear activation or is a consequence of it' [6]: from the present exposition it is clear that both physical separation of *Hox* genes and their collinear activation are indispensable and non-separable elements of a <u>single</u>

activation mechanism. This mechanism, based on the application of physical forces, underlies all molecular processes participating in the expression of clustered *Hox* genes [**A,B**]. The demonstration that *Hox* gene expression is tightly connected to fiber gene translocations supports the hypothesis of physical forces. After all, any movement from place to place is caused by the application of some force. Up to now, the idea of physical forces being involved in *Hox* gene collinearity has been persistently ignored in the established biomolecular models. However, the recent findings render this involvement unavoidable.

    5. It is surprising that a simplified mechanism like the biophysical model can satisfactorily explain such a wide range of phenomena and so complex experimental results. The speculation that physical principles might be involved (and in particular the Coulomb forces) probably reflects some inherent truth hidden in this model.

    6. Finally, the well known dialectic triad of reasoning (Thesis- Antithesis- Synthesis) could be applicable to the collinearity problem: the thesis (Physical Principles) and the antithesis (Chemical Principles) lead to the synthesis (Physicochemical cooperation). Ultimately, both models could participate in a natural synergy. The biophysical model fixing the temporal and spatial trigger and the biomolecular model (incorporating gene enhancers, inhibitors, *cis*-regulators etc) the mechanism of expression.

**References**

1. Lewis EB: **A gene complex controlling segmentation in *Drosophila*.**
   *Nature* 1978, **276:**565-570.

2. Lemons D, McGinnis W: **Genomic evolution of Hox gene clusters**.
   *Science* 2006, **313:**1918-1922.

3. Gehring WJ, Kloter U, Suga H: **Evolution of the Hox gene complex from an evolutionary ground state.**
   *Curr. Top. Dev. Biol.* 2009, **88:** 35-61.

4. Durston AJ, Jansen HJ, in der Rieden P, Hooiveld MHW: **Hox collinearity- a new perspective.** *Int. J. Dev. Biol.* 2011, **55:** 899-908.

5. Tschopp P, Tarchini B, Spitz F, Zakany J, Duboule D: **Uncoupling time and space in the collinear regulation of *Hox* genes.**
   *PloS Genetics* 2009, **5:**(3).

6. Noordermeer D, Leleu M, Splinter E. Rougemont J, de Laat W, Duboule D: **The

    **dynamic architecture of Hox gene clusters.**
    *Science* 2011, **334:** 222-225.

7. Tarchini B, Duboule D: **Control of *Hoxd* genes' collinearity during early limb development.**
    *Developmental Cell* 2006, **10:** 93-103.

8. Papageorgiou S: **A physical force may expose Hox genes to express in a morphogenetic density gradient.**
    *Bull. Math. Biol.* 2001, **63:** 185-200.

9. Papageorgiou S: **A cluster translocation model may explain the collinearity of Hox gene expressions.**
    *BioEssays* 2004, **26:** 189-195.

10. Papageorgiou S: **Pulling forces acting on *Hox* gene clusters cause expression collinearity.**
    *Int. J. Dev. Biol.* 2006, **50:** 301-308.

11. Papageorgiou S: **A biophysical mechanism may control the collinearity of *Hoxd* genes during the early phase of limb development.**
    *Hum. Genomics* 2009, **3:** 275-280.

12. Papageorgiou S: **Physical forces may cause Hox gene collinearity in the primary and secondary axes of the developing vertebrates.**
    *Develop. Growth & Differ.* 2011, **53:** 1-8. [**A**]

13. Papageorgiou S: **Comparison of models for the Collinearity of Hox genes in the developmental axes of vertebrates.**
    *Current Genomics* 2012, **13:** 245-251. [**B**]

14. Papantonis A, Cook PR: **Fixing the model for transcription: The DNA moves, not the polymerase.**
    *Transcription* 2011, **2:** 41-44.

15. Towers M, Wolpert L, Tickle C: **Gradients of signalling in the developing limb.**
    *Curr. Opin. Cell Biol.* 2011, **24:** 181-187.

**Fig.1**

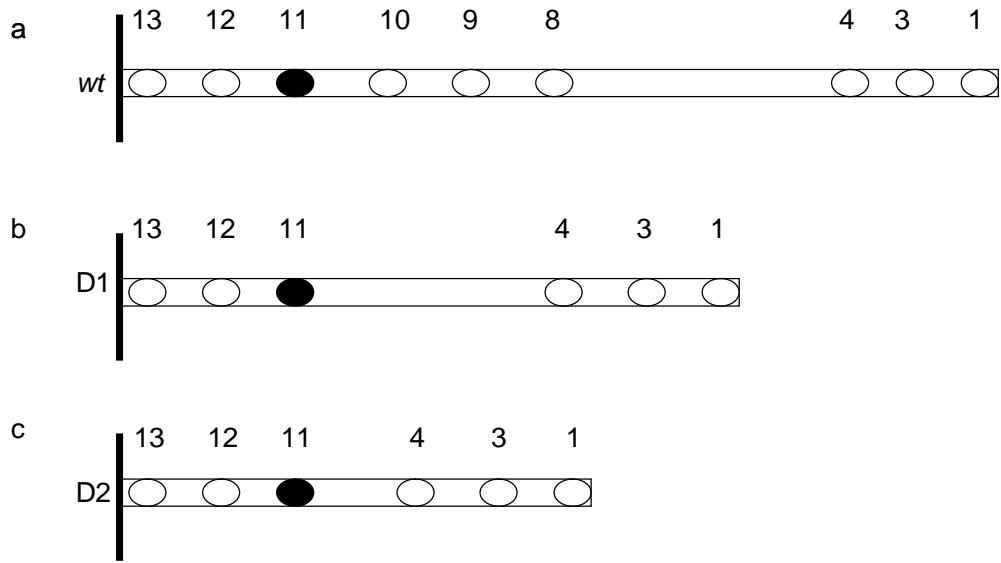

**Fig.2**

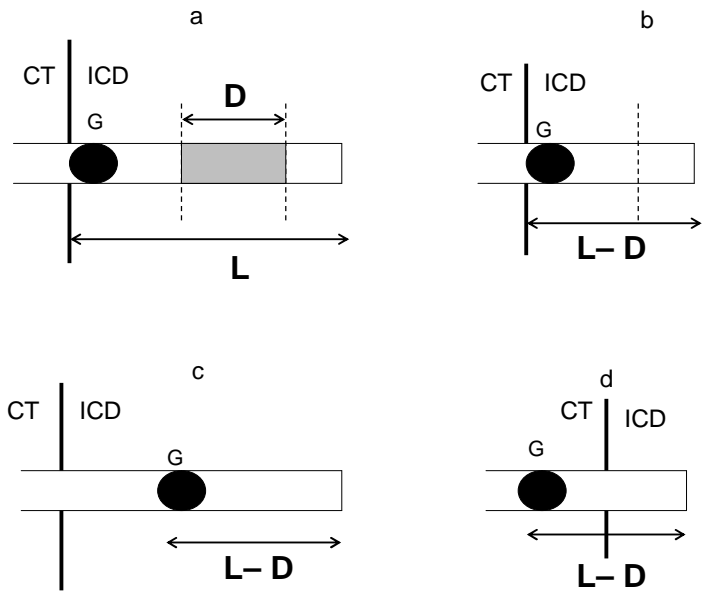

**Figure Captions**

**Fig. 1: Schematic representation of *HoxD* cluster and deleted DNA regions anterior to *Hoxd11* .**

a) The *wt HoxD* cluster and gene *Hoxd11* whose expression is probed. The length of the extruded fiber (from the anterior end to *Hoxd11*) is L.

b) The anterior region D1= [*Hoxd8-Hoxd10*] is deleted: del(8,10).

c) The anterior region D2= [i-*Hoxd8-Hoxd10*] is deleted: del(i,10). (i) is the intergenic region between *Hoxd4* and *Hoxd8*.

**Fig. 2: Expression of the mutant *Hox* gene G after anterior deletion D**

a) Mutant gene G moves from the Chromatin Territory (CT) where the gene is inactive toward the Interchromosome Domain (ICD) where the gene is activated. The extruded fiber length of the *wt* probe gene is L (from the anterior end of the cluster to G). D is an anterior DNA region to be deleted.

b) After the deletion of D, the extruded DNA length is $\underline{E = L-D.}$ In this case, the probe gene remains in the same position in ICD as in the *wt* case.

c) $\underline{E > L-D}$. The extruded length E exceeds L-D. The mutant gene G moves further inside ICD. For the activation of G, anteriorization is possible: G can retreat toward the interface CT/ICD.

d) $\underline{E < L-D}$. The mutant gene remains inside CT where its activation is impossible. G activation is possible only if G can move toward ICD (posteriorization).